\documentstyle[manuscript,prl,aps,epsfig]{revtex} 
\def\m{\mu}
\begin{document}
\draft
\preprint{MRI-PHY/P20000307} 

\title{\bf Are Messages of R-parity Violating Supersymmetry Hidden \\
within Top Quark Signals ?}
\author{Aseshkrishna Datta \footnote{E-mail: asesh@mri.ernet.in}
 and Biswarup Mukhopadhyaya \footnote{biswarup@mri.ernet.in}} 
\address{Mehta Research Institute, Chhatnag Road, Jhusi, Allahabad - 211 019, 
India}
\date{\today}
\maketitle
\begin{abstract}
In an R-parity nonconserving supersymmetric theory,
the lighter stop can dominantly decay into
$b\mu$ and $b\tau$ if R-parity breaking has to explain the
neutrino mass and mixing pattern suggested by the data on atmospheric muon 
neutrinos. This should give rise to $dilepton+dijet$ and
$single-lepton+jets$, signals identical with  those of the top quark at the
Fermilab Tevatron.
One can thus  constrain
the stop parameter space using the current top search data,
and similarly look for the first signals of supersymmetry at the upgraded 
runs of the Tevatron. 
\end{abstract}

%\begin{multicols}{2}
\narrowtext

It has been repeatedly suggested that in a supersymmetric (SUSY) scenario
\cite{habkane}, one of the two spin-zero superpartners
of the top quark, popularly known as stop, could be considerably lighter than 
any other strongly interacting superparticle \cite{lightstop}. 
This happens due to the mixing between the 
left-and right chiral stops, which can be quite large in principle, reducing
the mass of the lighter eigenstate.
Also, negative contribution from Yukawa coupling in the evolution
equations for scalar mass parameters can cause the stop to be 
lighter than the other squarks in a framework where all
squark flavours have the same mass at a high energy scale. It has been
shown that in the minimal supersymmetric standard model (MSSM), a light
stop ($\tilde{t}_1$) will decay dominantly through the channels 
$\tilde{t}_1 \to  c \chi^0_1$ and
$\tilde{t}_1 \to  b \chi^+_1$, where 
$\chi^0_1$ and $\chi^+_1$ are the lightest neutralino and the lighter chargino
respectively. Out of these, the former reigns supreme \cite{cneut} if the 
light stop is
even lighter than $\chi^+_1$ \cite{bwneut}. Based on the ensuing signals, 
experiments 
at the Fermilab Tevatron  have set a lower limit of
about $120-135$ GeV on the lighter stop \cite{stopbound} depending on the mass
of the lightest neutralino. 

The observed signals can be quite different when R-parity, defined as
$(-1)^{3B + L + 2S}$, is not a conserved quantity, something that is
both theoretically and phenomenologically consistent in supersymmetric
theories so long as only {\em one}  of  baryon number ($B$) and 
lepton number ($L$) is violated. In such cases, the lightest supersymmetric
particle (LSP) can be unstable, thereby altering the signals of SUSY
\cite{barger,godbole}. 
If, in addition, the LSP (which is the lightest neutralino in most theories)
is massive enough to evade searches at 
the large electron-positron collider (LEP), then the observation of SUSY 
at the Fermilab Tevatron may depend squarely on
the production and decays of the light stop for whom 
R-parity violation opens up new and often dominant 
decay modes. The limits based on  
$\tilde{t}_1 \to  c \chi^0_1$ require re-examination
in such a case.

In this letter, we want to point out that the easiest identifiable signatures
of a light stop (and therefore perhaps of SUSY itself) in an R-parity
nonconserving framework could be in final states that have been, and
still are, widely studied to look for the {\em top quark} at the Tevatron,
namely, the $dimuon+dijet$ as well as $single~muon+jets$ signals. 
We further argue that stop decays leading to such final states are
imperative in R-parity violating scenarios that lead to neutrino
masses and mixing as required by the recent SuperKamiokande (SK) data
on atmospheric muon neutrinos. Thus a careful
reanalysis of the already existing Tevatron Run I results on top search
may allow us to explore a large region of the light stop 
parameter space of such theories, where the search channel employed for MSSM
is not going to be effective.  
In Run II one can have an even wider prospect, not only
for constraining the SUSY parameter space but also for the possibility
of top signals being actually faked by the stop. In addition,
some new signals for such models, very similar to those of 
the top quark, are suggested here.

Expressed in terms of the quark, lepton and Higgs superfields, the MSSM 
superpotential is

%\begin{equation}
\begin{eqnarray}
W_{MSSM} &=& {\mu} {\hat H}_1 {\hat H}_2 + h_{ij}^l {\hat L}_i {\hat
H}_1 {\hat E}_j^c
+ h_{ij}^d {\hat Q}_i {\hat H}_1 {\hat D}_j^c  %\nonumber \\* &&
+ h_{ij}^u {\hat Q}_i {\hat H}_2 {\hat U}_j^c
\end{eqnarray}
%\end{equation}
where $\m$ is the Higgsino mass parameter and the last three terms give all the
Yukawa interactions.

The possible additions to this superpotential due to R-parity violation 
(through lepton number violation only) are given by \cite{barger}

\begin{equation}
W_{\not L} =  \epsilon_i {\hat L}_i {\hat H}_2 +            
\lambda_{ijk} {\hat L}_i {\hat L}_j {\hat E}_k^c +
\lambda_{ijk}' {\hat L}_i {\hat Q}_j {\hat D}_k^c 
\end{equation}  

\noindent
We consider only the effects of the trilinear additional terms in the
superpotential; we shall point out at the end of the paper that the effects
we are suggesting can arise also from the bilinear terms
$\epsilon_i {\hat L}_i {\hat H}_2$. 

If the SK data \cite{fukuda} on atmospheric muon neutrinos 
(and also similar data from the
Soudan-II \cite{soudan2} and MACRO \cite{macro} experiments)  
have to be explained in terms
of $\nu_\mu - \nu_\tau$ oscillations,
then the mass-squared splitting between the second and third lightest physical 
neutrino states will have to be $\Delta m^2_{23} \simeq 5 \times 10^{-3}$,
in addition to near-maximal mixing between the corresponding flavour 
eigenstates. R-parity violation in the form of the 
$\lambda$-and  $\lambda^{'}$-type interactions in 
equation 2 can give rise to neutrino
mass terms at one-loop level, the generic expression for them (in the 
flavour basis) being 
\begin{eqnarray}
(m^{\rm loop}_\nu)_{ij} &\simeq& \frac {3} {8\pi^2}  m^d_k m^d_p M_{SUSY}
\frac {1} {m^2_{\tilde q}} {\lambda_{ikp}'\lambda_{jpk}'}
\nonumber\\* && +
\frac {1} {8\pi^2}  m^l_k m^l_p M_{SUSY}
\frac {1} {m^2_{\tilde l}} {\lambda_{ikp}\lambda_{jpk}}
\end{eqnarray}
where $m^{d(l)}$ denote the down-type quark (charged lepton) masses.
${m^2_{\tilde l}}$, ${m^2_{\tilde q}}$ are the average slepton and
squark mass squared. $M_{SUSY}(\sim \mu)$ is the effective scale of
supersymmetry breaking. The mass and mixing patterns 
suggested by the observed $\nu_\mu$ data can be accommodated
in the above scheme of mass generation if,
with $M_{SUSY} \simeq m{\tilde{q}} \simeq 300-500$ GeV, 
$\lambda^{'}_{233} \simeq \lambda^{'}_{333} \simeq ~a~few~times~ 10^{-4}$
\cite{bakira}.
An immediate consequence of such couplings is the decay of the lighter 
stop in the channels
$\tilde{t}_1 \to b \tau^{+}$ and 
$\tilde{t}_1 \to b \mu^{+}$ with comparable widths.

There is a considerably large region of the parameter space, not yet 
constrained by any experimental data, where the stop is the second lightest 
supersymmetric particle next to the lightest neutralino. In this region,
the three lowest-order decay channels available to the stop are
$\tilde{t}_1 \to  c \chi^0_1$,
$\tilde{t}_1 \to b \tau^{+}$ and 
$\tilde{t}_1 \to b \mu^{+}$ \cite{diazvalle}. The first one, a well-studied
process, is a consequence of neutral flavour violation in SUSY and is 
suppressed by the small mismatch between quark and squark mass matrices. The
latter ones are driven by $\lambda^\prime_{233}$ and $\lambda^\prime_{333}$,
and we find that they dominate for a light stop
with mass $\le 150$ GeV so long as 
$\lambda^{'}_{233},\lambda^{'}_{333}$ lie in the range specified above,
in conformity with the SK data. 

We focus our attention on single-muon and dimuon 
final states, together
with jets and missing energy, arising from (light) stops
pair-produced in $p{\overline p}$ collisions at the Fermilab Tevatron,
with either both decaying into $b\tau$ or one of them into $b\tau$ 
while the other goes to $b\mu$.   
The same final states have been extensively analysed for the determination
of the mass and production cross-section of the top quark. 
It should be noted that the $\tau$  produced together with a $b$ quark in the 
two-body decay of a stop (with mass $100-150$ GeV) 
can have sufficient $p_T$ for the
resulting jets/muons/neutrinos to often pass the $p_T$ or $\not{E_T}$ cuts
associated with the top quark signals. Thus, depending upon whether
the tau decays hadronically or semileptonically, the stop decays can 
contribute to top-like signals of the form $(i)~single~muon+3jets+\not{E_T}$
and $(ii)~dimuon+2jets+\not{E_T}$, provided that the jets/leptons satisfy
the requisite cuts.

We present some numerical estimates where the five degenerate squark flavours 
are assumed to have masses $\approx 400$ GeV. The mass of the lighter stop is 
made to vary beween $80$ and $150$ GeV, while the $SU(2)$
gaugino mass $M_2$ is held fixed at $150$ GeV. Unification of
the gaugino masses has been assumed. In addition, $\tan\beta$, 
the ratio of the 
two Higgs expectation values, has been fixed at 3. Although we have \emph{not}
assumed any definite high scale SUSY breaking mechanism, it can be verified 
that such combinations of parameters can indeed be realised in a
supergravity (SUSY) framework. This is possible by using one's freedom 
\cite{dboer} with 
the trilinear soft SUSY breaking term $A$ (while still preserving charge and 
colour invariance \cite{casas}) and using the Higgsino parameter 
$\mu$ as a phenomenological
input \cite{manuel}, something that is justfied if the Higgs  
mass parameters retain the
freedom of differing from the `universal' sfermion mass parameter ($m_0$) at
the grand unification scale.

The above specified set of parameters allows one to calculate the R-conserving 
stop decay width. For the R-parity violating two-body decays, we have 
used three sets of values for $\lambda^{'}_{233}=\lambda^{'}_{333}$, 
consistent with the expectation from the SK results. Using these,
it is straightforward to calculate the branching ratios of the various stop decay
modes. It is found that for the region of parameter space under investigation
here, the branching ratio for R-parity violating decays ranges from
75 to 99 per cent, with a near-equal
share between the $b \tau$ and $b \mu$ channels.

A parton level Monte Carlo calculation has been performed 
for both $single~muon+three~jets+\not{E_T}$ and 
$dimuon+two~jets+\not{E_T}$ final states arising from stop pair-production
at the Tevatron. In the former case, we have demanded that at least one 
jet be identified as a $b$-induced one. Both the top- and stop-production
cross-sections have been QCD corrected using a a so-called $K$-factor of
1.4 in case of top \cite{kfactortop} and 1.3 \cite{kfactorstop} for the stop. 
For hadronic decays of the tau, we have considered only modes with one
charged track, arising from $\tau^\pm \to \pi^\pm \nu_{\tau}, 
~\rho^\pm \nu_{\tau}, ~a_1^\pm \nu_{\tau}$. 
In order to calibrate our parton-level
results, we have computed the numbers of the same types of expected 
signal events from top quark pair production. The numbers of 
both single-muon and dimuon events
thus calculated in the latter case agree, within small errors, with
the actual observations \cite{cdf1,cdf2}, so that our results may be treated 
as reasonable estimates of how many stop-induced events can be contained in the
top signal.

Since our main purpose is to see how stop signal can percolate into top signals
for which detectors are already designed, both have been subjected to the 
same set of cuts, as specified by the CDF top search strategy. 
In both types of final states of our interest, jets are defined using the 
cone algorithm as discussed in Ref.\cite{cdf1,cdf2}. Jets are counted in the 
analysis only if $|\eta |_{jet} < 2$ where 
$\eta$ is the  pseudorapidity \cite{cdf1,cdf2}. 
For single muon final state we require jets with $E_T > 15$ GeV, at least 
one of which is identified as a $b$-jet (with identification
efficiency $\approx 40$ \% \cite{cdf1}) while for dimuon final state two jets 
with $E_T > 10$ GeV are required \cite{cdf2}. Two jets are merged if 
$\Delta R \le 0.4$, where $\Delta R ~=~\sqrt{\Delta \eta^2 + \Delta \phi^2}$,
$\Delta \eta$ and $\Delta \phi$ being the separations in pseudorapidity
and azimuthal angle. 
In both the cases \emph{isolated} 
muon(s) with $p_T > 20$ GeV are necessary in the central region with $|\eta |
< 1$ \cite{cdf1,cdf2}. The criterion for muon isolation imposed here is that
the total $E_T$ within a cone of $\Delta R~=~0.4$ around it should 
be less than 10 \% of its own $p_T$.  
In dimuon final states, backgrounds from real $Z$ are 
rejected by requiring the dimuon invariant mass to be outside 
the interval $75-105$ GeV \cite{cdf2}. For single muon events, 
missing transverse
energy, $\not E_T > 20$ GeV is demanded \cite{cdf1} while for dimuons
$\not E_T > 25$ GeV is generally required \cite{cdf2}. It is observed in the
last of Ref. \cite{cdf2}  
that this combination of cuts effectively eliminates backgrounds
from Drell-Yan production and other relevant processes for the
dimuon final state. 

The net contribution of the stop to the top-like signals depends on two
effects: the production cross section of a stop pair vs that of a top pair,
and the relatively large branching fraction for the stops cascading into
the final states of our interest. In addition, of course, the susceptibility
to cuts plays a role. For $m_{\tilde{t}_1} \approx 100$ GeV the production
rates approximately match, and the stop production rate falls for higher
stop masses.

Figures 1 and 2 contain our main results, based on an integrated luminosity
of $109$ pb$^{-1}$. We have checked by varying the 
parameter  $\mu$ between $-700$ and  $+700$ GeV that the results are
not altered by more than about 25 per cent. 
The dimuon events enable us to rule out a slightly larger range of the stop 
mass. This is because of more suppression of the top signal in this channel
through leptonic branching fractions of $W$ as opposed to what happens for
stop-pairs. It is clear from
figure 2 that the dimuon data should definitely rule out the entire stop mass
interval that is otherwise constrained assuming 
$B(\tilde{t}_1 \to c \chi^0_1) \simeq 1$. In fact, the
limit can perhaps be pushed somewhat higher up in the R-parity violating case.
For $\lambda^{'}_{233} \simeq \lambda^{'}_{333}=10^{-4}$, 
this limit should be around 125 GeV with the currently available data, 
while for $\lambda^{'}$-values of $5 \times 10^{-4}$ it 
touches about 140 GeV.

From the single muon signal too, stop masses upto about $120-125$ GeV 
seem to be ruled out. We also want to emphasize that both this channel
and the dimuon one can be used effectively in the context of Run II
of the Tevatron to look for SUSY signals burried within the top data. A careful
analysis of the dimuon versus dielectron data there, invigorated by the 
additional available luminosity, may lead to successful identification of an
excess in the former, thereby indicating R-parity violation of a kind
that simultaneously explains the atmospheric neutrino puzzle.

For simplicity, we have confined ourselves in the above discussion to
cases where the stop is kinematically forbidden to decay into a $b$ quark and a 
chargino. Such a decay can be possible when $M_2$ and consequently the 
mass of lightest 
neutralino is smaller. However, in view of the fact that the chargino can 
subsequently decay into a lepton, the above mode can
in effect strengthen the stop contribution.

The values of R-parity violating couplings which are as small as the ones under
study here are unlikely to contribute to precision electroweak data 
\cite{erler} or
to R-parity violating decays of the top quark \cite{ag} such as 
$t \longrightarrow b \tilde{l}$ or $t \longrightarrow \tilde{b} l$ even if they
are kinematically allowed. However, with couplings several times larger in 
magnitude (with larger average squark mass, for example),
some of these decays could lead to additional 
signals (final states of higher multiplicity) from the decays of the top,
particularly at Run II of the Fermilab Tevatron.

Another signal, similar in nature to the ones discussed above, is
one where both the stops decay into the $b \mu$ channel. In this case one
shall see dimuons with two jets. Proper $b$-identifaction in the jets, 
together with effective measures to eliminate the Drell-Yan backgrounds,
can establish such final states as rather effective ones for discovering
a light stop in an R-parity violating theory. 

As has been mentioned at the beginning, the above discussion has assumed
only trilinear R-parity violating interactions. The presence of the
bilinear terms of the form $L_{i}H_2$ in the superpotential (and the 
consequent vacuum expectation values for sneutrinos) can give rise
to mixing between charged leptons and charginos \cite{blrpv}, 
which can again lead to  
top-like signals of the same kinds from stop decays. A detailed analysis
of the signals in such a case will soon be presented by us \cite{future}.

In conclusion, the first signals of  R-parity violating
SUSY observable in current and upcoming experiments 
can very well mimic those of the top quark. This is
particularly true if the theory has to account for 
the neutrino masses  and mixing  suggested
by experimental results on atmospheric muon neutrinos. We can 
use the already available $single~lepton+jets$ as well as 
\emph{dilepton} data from the Fermilab Tevatron to constrain a large  
range of the lighter stop mass in such a scenario. In the 
upgraded version of the Tevatron,
too, top quark  signals will continue to be useful probes for such 
types of SUSY. 
  
We thank S. Raychaudhuri and M. Guchait for technical help.

\begin{figure}[htb]
\vspace*{-1.00in}
\centerline{
\epsfig{file=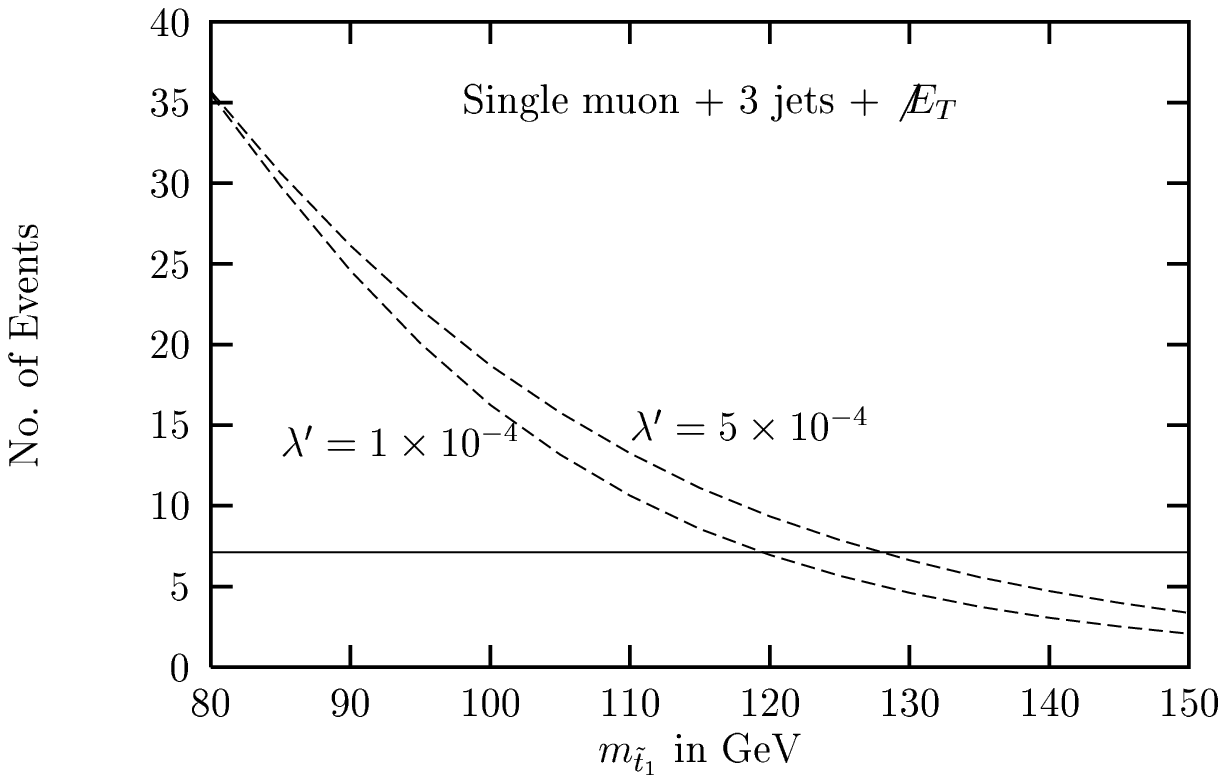, height=20cm,width=15cm}
}
\vspace{-4.3in}
%\vspace{-3.3in}
\caption{Contributions to single muon + 3 jets + $\not E_T$ events from 
$t\bar{t}$ production for $m_t=175$ GeV (solid line) and 
$\tilde{t}_1 \tilde{t}_1^*$ production
(followed by R-parity violating decays of $\tilde{t}_1$'s, dashed lines) 
as a function of $m_{\tilde{t}_1}$ at Tevatron with $\sqrt{s} = 1.8$ TeV 
and an integrated luminosity of 109 pb$^{-1}$ and using the same set of cuts
(see text). 
The MSSM parameters used are : 
$M_2=150$ GeV, $\mu=~-400$ GeV, $\tan \beta =3$, $\theta_{\tilde{t}} ({\rm
stop~mixing~angle}) \approx -45^\circ$, $m_{Q,U,D}=400$ GeV for first 
two generations
of squarks. We take $\lambda^\prime = \lambda^\prime_{233} = 
\lambda^\prime_{333}$.
CTEQ-4M parton distributions are used.}
\end{figure}

\begin{figure}[htb]
\vspace*{-1.00in}
\centerline{
\epsfig{file=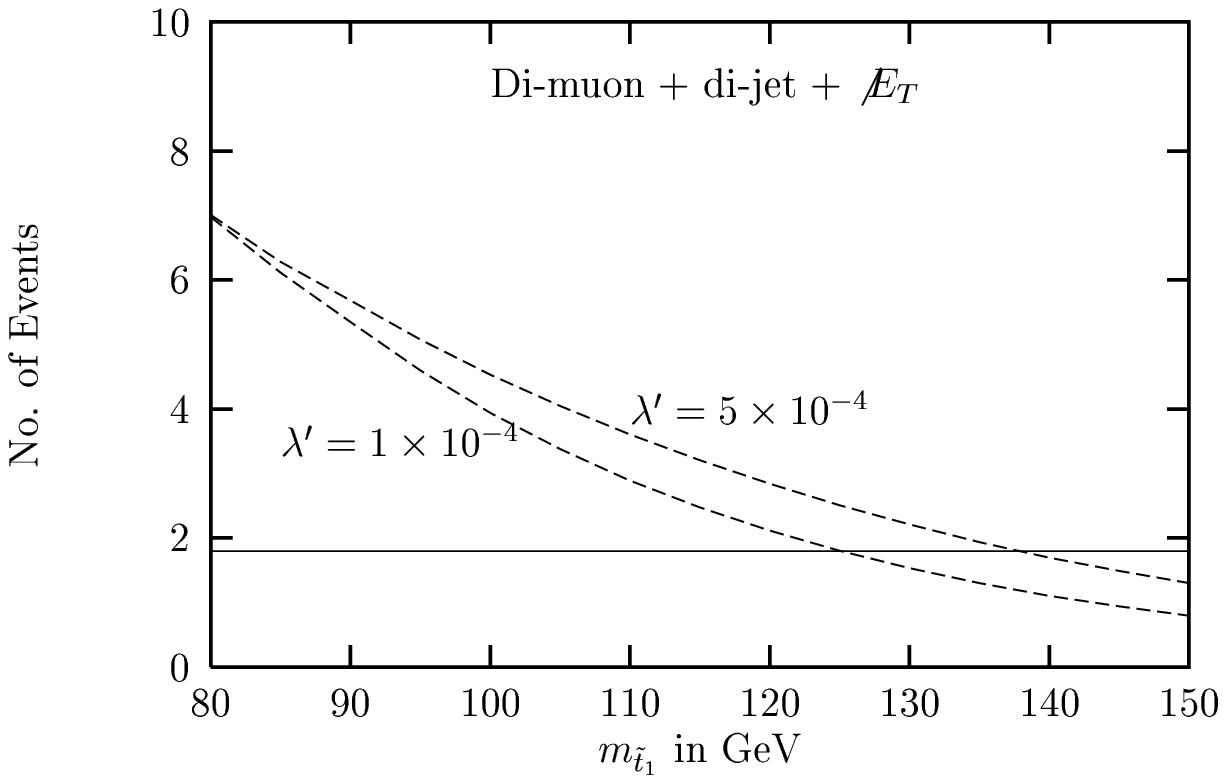, height=20cm,width=15cm}
}
\vspace{-4.3in}
%\vspace{-3.3in}
\caption{Same as in Fig.~1 except for contributions to di-muon + di-jet + 
$\not E_T$ events.}
\end{figure}

%\end{multicols}


\begin{thebibliography}{99}

\bibitem{habkane}  For reviews, see, e.g., 
H.P. Nilles, Phys. Rep. {\bf 110}, 1 (1984); 
H.E. Haber and G.L. Kane, Phys.
Rep.  {\bf 117}, 75 (1985); G. Kane(ed.), Perspectives on
Supersymmetry (World Scientific).

\bibitem{lightstop} J. Ellis  and S. Rudaz, Phys. Lett. B {\bf 128}, 248 
(1983). 

\bibitem{cneut}
K. Hikasa and M. Kobayashi, Phys. Rev. D {\bf 36}, 724 (1987);
H. Baer et al., Phys. Rev. D {\bf 44}, 725 (1991);
H. Baer, J. Sender and X. Tata, Phys. Rev. D {\bf 50}, 4517 (1994).

\bibitem{bwneut}
In addition, three-body decays of the stop can be of some significance over a
limited region of the parameter sapce. For discussions see, e.g., 
A. Bartl et al., Z. Phys. C {\bf 73}, 469 (1997); W. Porod and T. W\"{o}hrmann,
Phys. Rev. D {\bf 55}, 2907 (1997); W. Porod, Phys. Rev. D {\bf 59}, 095009
(1999); A. Datta, M. Guchait and K.K. Jeong, Int. Jour. of Mod. Phys. 
A {\bf 14} 2239 (1999).

\bibitem{stopbound}
CDF Collaboration, F. Abe et al., Phys. Rev. Lett. {\bf 83}, 2133 (1999);
CDF Collaboration, T. Affolder et al., hep-ex/9910049.
 
\bibitem{barger} 
V. Barger, G. Giudice and T. Han, Phys. Rev. D {\bf 40}, 2987 (1989);

\bibitem{godbole}
D.P. Roy, Phys. Lett. B {\bf 283}, 270 (1992); 
D.K. Ghosh, S. Raychaudhuri and K. Sridhar, Phys. Lett. B {\bf 396}, 177 
(1997); 
R.M. Godbole, P. Roy and X. Tata, Nucl. Phys. B {\bf 401}, 67 (1993);
D. Chowdhury and S. Raychaudhuri, Phys. Lett. B {\bf 401}, 54 (1997);
H. Dreiner, hep-ph/9707435, published in \emph{Perspective on Supersymmetry}, 
ed. by G.L. Kane; 
\emph{Report of the Group on R-parity Violation}, R. Barbier et al.,
hep-ph/9810232. 

\bibitem{fukuda}
SuperKamiokande Collaboration, Y. Fukuda et. al., Phys. Rev.Lett. {\bf 81}, 
1562 (1998).

\bibitem{soudan2}
SOUDAN2 Collaboration,
W.W.M. Allison et al., Phys. Lett. B {\bf 449}, 137 (1999).

\bibitem{macro} MACRO Collaboration,
M. Ambrosio et al., Phys. Lett. B {\bf 434}, 451 (1998).

\bibitem{bakira} M. Drees, S. Pakvasa, X. Tata, and T. ter Veldhuis, Phys.
Rev. D {\bf 57}, 5335 (1998); 
B. Mukhopadhyaya, S. Roy, and F. Vissani, Phys. Lett. B {\bf 443}, 191 (1998).
S. Rakshit, G. Bhattarchyya, and A.  Raychaudhuri, Phys. Rev.
D {\bf 59}, 091701 (1999).

\bibitem{diazvalle}
H. Dreiner and R.J.N. Phillips, Nucl. Phys. B {\bf 367}, 591 (1991); 
F. de Campos et al., hep-ph/9903245;
M.A. D\'\i az et al., hep-ph/9908286;
M.A. D\'\i az, hep-ph/9911274;
W. Porod, D. Restrepo and J.W.F. Valle, hep-ph/0001033.

\bibitem{dboer} 
W. de Boer, R. Ehret and D.I. Kazakov, Z. Phys. C {\bf 67}, 647 (1995).

\bibitem{casas}
J.A. Casas, A. Lleyda and C. Munoz, Nucl. Phys. B {\bf 47}, 3 (1996).

\bibitem{manuel}
C. Boehm, A. Djouadi and M. Drees, hep-ph/9911496.

\bibitem{kfactortop}
E. Berger and H. Contopanagos, Phys. Rev. D {\bf 54}, 3085 (1996);
S. Catani, M. Mangano, P. Nason and L. Trentadue, Phys. Lett. B {\bf 378},
329 (1996).

\bibitem{kfactorstop}
W. Beenakker et al., Nucl. Phys. B {\bf 515}, 3 (1998).

\bibitem{cdf1} 
CDF Collaboration, F. Abe et al., Phys. Rev. Lett. {\bf 74}, 2626 (1995);
CDF Collaboration, F. Abe et al., Phys. Rev. Lett. {\bf 80}, 2773 (1998).


\bibitem{cdf2}
CDF Collaboration, F. Abe et al., Phys. Rev. Lett. {\bf 80}, 2779 (1998);
CDF Collaboration, F. Abe et al., Phys. Rev. Lett. {\bf 82}, 271 (1999).


\bibitem{erler} J. Erler, J. Feng and N. Polonsky, Phys. Rev. Lett. 
{\bf 78}, 3063 (1997).

\bibitem{ag} K. Agashe and M. Graesser, Phys. Rev. D {\bf 54}, 4445 (1996); 
L. Navarro, W. Porod and J. Valle, Phys. Lett. B {\bf 459}, 615 (1999).


\bibitem{blrpv} L. Hall and M. Suzuki, Nucl. Phys. B {\bf 231}, 419 (1984);
R. Hempfling, Nucl. Phys. B {\bf 478}, 3 (1996);
H. P. Nilles and N. Polonsky, Nucl. Phys. B {\bf 484}, 33 (1997);
S. Roy and B. Mukhopadhyaya, Phys. Rev. D {\bf 55}, 7020 (1997);
M.A. Garc{\'\i}a-Jare\~no, and J. W. F. Valle, Nucl. Phys. B {\bf 529}, 
3 (1998);
M.A. D\'\i az, J. Ferrandis, J.C. Rom\~ao, and J.W.F. Valle, 
Phys. Lett. B {\bf 453}, 263 (1999);
M.A. D\'\i az, E.Torrente-Lujan, J.W.F. Valle, 
Nucl. Phys. B {\bf 551}, 78 (1999); 
E.J. Chun, S.K. Kang, C.W. Kim, and U.W. Lee, Nucl. Phys. B {\bf 544}, 89 
(1999); 
A. S. Joshipura and S.K. Vempati, Phys. Rev. D {\bf 60}, 095009 (1999);
Chao-Hsi Chang and Tai-Fu Feng, Eur. Phys. Jour., C {\bf 12}, 137 (2000);
A. Datta, B. Mukhopadhyaya and S. Roy, Phys. Rev. D {\bf 61}, 055006 (2000).

\bibitem{future}
A. Datta and B. Mukhopadhyaya, in preparation.
\end{thebibliography}
\end{document}